\documentclass[conference]{IEEEtran}
\usepackage{booktabs}
\usepackage{multirow}
\IEEEoverridecommandlockouts
\usepackage{amsmath,amssymb,amsfonts}
\usepackage{algorithmic}
\usepackage{graphicx}
\usepackage{textcomp}
\usepackage{xcolor}
\usepackage[backend=bibtex,bibstyle=ieee,citestyle=numeric-comp]{biblatex}
\usepackage{caption} 
\usepackage{hyperref}
\def \partha [#1]{\textcolor{red}{Partha: #1}} 

\begin{document}

\title{Self-Supervised Object-Centric Representation for Multi-Task Audio Analysis \\
\thanks{The funding for this work is supported by Jane and Aatos
Erkko Foundation through the CONVERGENCE of Humans
and Machines project.}
}


\author{
    \IEEEauthorblockN{ Parthasaarathy Sudarsanam, Tuomas Virtanen}
    \IEEEauthorblockA{\textit{Audio Research Group, Tampere University, Finland}}
    \IEEEauthorblockA{\{parthasaarathy.ariyakulamsudarsanam, tuomas.virtanen\}@tuni.fi}
}

\maketitle
\begin{abstract}
This paper studies the capability of an object-centric audio representation to support multiple downstream tasks within a unified framework. The object-centric model is trained in a self-supervised setup using slot attention by reconstructing Audio Spectrogram Transformer (AST) patch embeddings from a set of latent slots. It is trained on synthetic audio mixtures, which provides controlled multi-source audio scenes. These slot representations are evaluated for the classification, detection, and separation of sound events by training task-specific output layers. On the classification task, the model achieves an F1-score of 0.8, comparable to clip-level AST embeddings which achieve 0.7. For sound event detection, it reaches a macro F1-score of 0.7, outperforming clip-level AST embeddings at 0.4 and matching AST patch embeddings at 0.7, which encode finer time–frequency information. In source separation, it achieves an SI-SDR improvement (SI-SDRi) of 18.4\,dB. Furthermore, in a zero-shot source separation setting, the model achieves an SI-SDRi of 14.5\,dB. These results demonstrate that slot-based representations capture transferable and compositional structure in audio scenes, enabling generalization across tasks and datasets.
\end{abstract}


\section{Introduction}


Understanding complex acoustic scenes requires representations that can disentangle and organize multiple simultaneous sound sources. Humans can naturally focus on a sound of interest within a complex scene, selectively attending to one source while ignoring others, depending on the context. This suggests that representing audio as a set of distinct components can better support tasks such as sound event detection and source separation. However, many existing audio representation learning models such as CLAP \cite{elizalde2023clap}, PANNs \cite{kong2020panns}, VGGish \cite{hershey2017cnn},  encode acoustic scenes into clip-level embeddings without explicitly modeling individual sources. As a result, overlapping events, such as speech, music, and environmental sounds, become entangled, making source-level classification, temporal localization, and source separation more difficult.

Recent advances in object-centric learning suggest that structured latent representations can provide a more interpretable and compositional alternative. In particular, slot-based models have been proposed as a mechanism to decompose input into a set of latent slots \cite{locatello2020object}, each intended to represent a distinct object or sound event in the scene. Although this idea has been extensively explored in the image domain for object discovery and scene decomposition \cite{locatello2020object, seitzer2023bridging, NEURIPS2022_735c847a, yuan2023compositional} , its application to audio remains limited, despite the inherently compositional nature of sound mixtures.

\begin{figure*}[t]
    \centering
    \includegraphics[width=\textwidth]{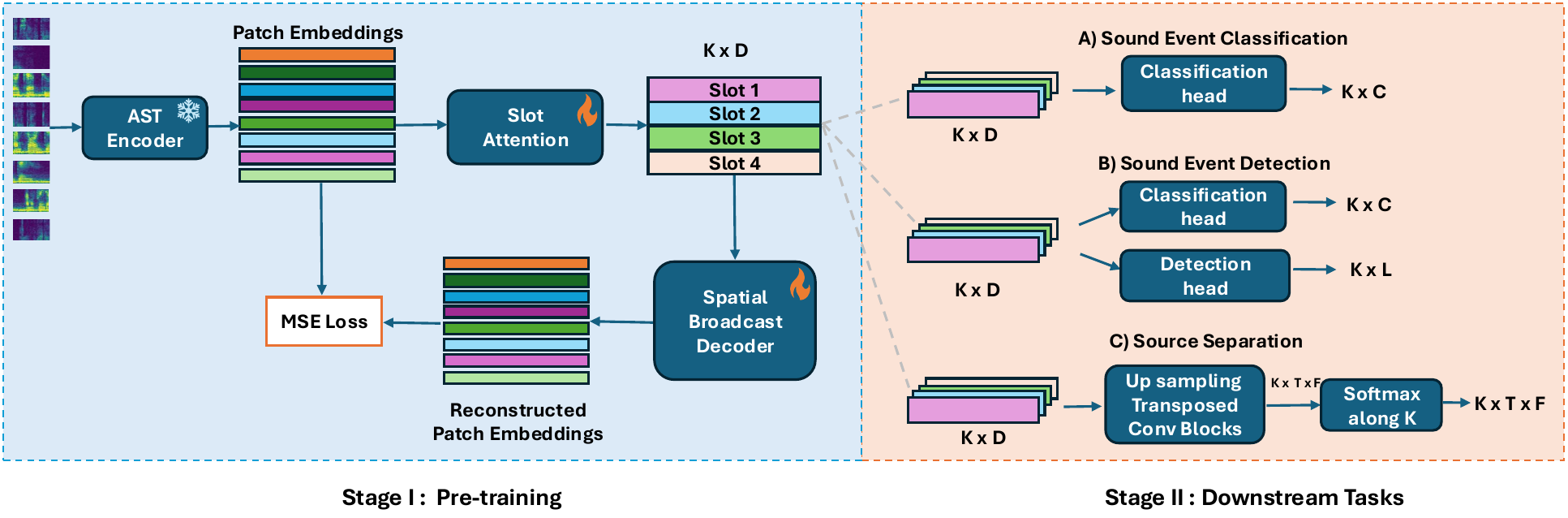}
    \caption{\textbf{Stage I}: Self-supervised slot pretraining. \textbf{Stage II}: Downstream classification, detection, and source separation tasks.}
    \label{fig:slot_architecture}
\end{figure*}

Some works have begun exploring object-centric representations for audio \cite{reddy2023audioslots, gha2023unsupervised, sridhar2025compositional, lee2026deepasa}, decomposing acoustic scenes into their constituent sound sources. In \cite{reddy2023audioslots}, the authors proposed a strongly supervised source separation model that learns source-specific representations using slot attention, demonstrating  source separation capabilities. In \cite{sridhar2025compositional}, the authors investigated compositional audio representations using slot attention and provide a comprehensive study exploring various design choices, including supervised and unsupervised training regimes as well as different reconstruction targets such as feature embeddings and spectrograms. Their experiments primarily focus on classification, while the potential of such representations for temporal detection and source separation, remains an open direction for exploration. In \cite{lee2026deepasa}, the authors introduced DeepASA, an object-centric framework for spatial audio that employs a convolutional object splitting module to decompose scenes into object streams, trained in a fully supervised manner using task-specific decoders for audio reconstruction, sound event detection, and direction-of-arrival estimation.

Although these works demonstrate the promise of object-centric representations for audio, several important challenges remain. Existing methods rely on strong supervision \cite{reddy2023audioslots, lee2026deepasa} or evaluate the learned representations on a limited set of downstream tasks \cite{sridhar2025compositional}. In addition, zero-shot capabilities of such representations have not been explored. These gaps leave open the question of whether object-centric audio representations can be learned in a fully self-supervised manner and reused across diverse downstream tasks.

In this work, we investigate the capabilities of a self-supervised object-centric audio representation learning framework. We show that the resulting slot representations generalize effectively across multiple downstream tasks and zero-shot settings, enabling a shared representation space for classification, detection, and source separation. Specifically, we make the following contributions:

\begin{itemize}
    \item Demonstrate that a single shared slot space can generalize to multiple downstream tasks, including the classification, detection, and source separation.

    \item Show that slot representations enable strong zero-shot source separation on unseen sound events.
\end{itemize}

\section{METHODS}\label{methods}

Our framework consists of two stages as shown in \ref{fig:slot_architecture}. A self-supervised object-centric representation pretraining stage and a downstream task adaptation stage. An Audio Spectrogram Transformer (AST) \cite{gong21b_interspeech} encodes the input audio spectrogram patches into patch embeddings, which are processed by a slot attention module to produce a set of latent slots. A decoder reconstructs these patch embeddings from the slots during pretraining.

After pretraining, the encoder and slot attention are frozen, and lightweight task-specific heads are trained for classification, detection, and source separation. 

\subsection{Self-Supervised Slot Representation Learning} \label{SS-SRL}

\textbf{Encoder:} Given an input audio, we first extract patch-level representations using a pretrained AST. The waveform is converted into a log-mel spectrogram using the AST feature extractor. The spectrogram is then divided into overlapping patches of size $16 \times 16$ with a stride of $10 \times 10$, and passed through the AST encoder to obtain a sequence of embeddings $\mathbf{X} \in \mathbb{R}^{N \times D}$, where $N$ denotes the number of patches and $D$ the embedding dimension. The encoder is kept frozen during training, and we discard the classification and distillation tokens, retaining only patch embeddings.

\textbf{Slot-Attention:} The resulting patch embeddings $\mathbf{X}$ are processed by a slot attention module, implemented using an inverted attention transformer \cite{wu2023inverted}, that maps them to a fixed set of $K$ latent slots $\mathbf{S} \in \mathbb{R}^{K \times D}$. We use $K=4$ slots and perform $T=3$ iterative refinement steps. The slots are initialized at $t=0$ using random values drawn from a learnable Gaussian distribution parameterized by $\boldsymbol{\mu}$ and $\boldsymbol{\sigma}$:

\begin{equation}
\mathbf{S}^{(0)} = \boldsymbol{\mu} + \boldsymbol{\sigma} \odot \epsilon, \quad \epsilon \sim \mathcal{N}(0, I)
\end{equation}

At each iteration $t$, we compute queries, keys, and values using learnable linear mappings $f_q, f_k$ and $f_v$:
\begin{equation}
\mathbf{Q} = f_q(\mathbf{S}^{(t)}), \quad
\mathbf{K} = f_k(\mathbf{X}), \quad
\mathbf{V} = f_v(\mathbf{X})
\end{equation}

We use inverted attention to compute slot-to-patch assignments, where the softmax is applied over slots, inducing competition among slots for each input patch. 
\begin{equation}
a_{k,n} = \mathrm{Softmax}_k \left( \frac{\mathbf{Q}_k \mathbf{K}_n^\top}{\sqrt{D}} \right)
\end{equation}

\noindent where, k indexes the slots and n indexes the input patches. Slot updates for each iteration are computed as
\begin{equation}
\Delta \mathbf{S}_k = \sum_{n=1}^{N} a_{k,n} \mathbf{V}_n
\end{equation}

Each slot is refined using a residual connection that adds the attention-based update and a multi-layer perceptron (MLP) applied after layer normalization (LN):

\begin{equation}
\mathbf{S}_k^{(t+1)} = \mathbf{S}_k^{(t)} + \Delta \mathbf{S}_k + \mathrm{MLP}(\mathrm{LN}(\mathbf{S}_k^{(t)}))
\end{equation}

This iterative process encourages a soft partitioning of the input representation, where each slot specializes in modeling a distinct subset of the acoustic scene.

\textbf{Decoder:} To reconstruct the patch embeddings, we use a spatial broadcast decoder \cite{seitzer2023bridging} that maps slot embeddings back to patch-level features. Given slot embeddings $\mathbf{S} \in \mathbb{R}^{K \times D}$, each slot embedding is repeated for every patch location and concatenated with learnable positional embeddings corresponding to each location. The resulting slot--position features are processed by a shared MLP that predicts a slot-specific patch embedding together with an additional scalar per slot and patch, used as a mixing score. A softmax is applied over slots at each patch location to normalize these scores, yielding slot assignment weights that determine the contribution of each slot to each patch embedding. The final reconstructed patch embeddings $\hat{\mathbf{X}} \in \mathbb{R}^{N \times D}$ are obtained as a weighted combination of slot-specific patch predictions. Implementation details can be found in the codebase\footnote{https://github.com/partha2409/Object-Centric-Audio-Representation \label{footnote:code}}.

The model is trained using a mean-squared error reconstruction loss between the predicted embeddings $\hat{\mathbf{X}}$ and the original encoder embeddings $\mathbf{X}$. This objective encourages the slots to encode complementary information that explains the full mixture. We perform this self-supervised pretraining for 200k steps using the Adam optimizer with a learning rate of $1\times10^{-4}$ and a weight decay of $1\times10^{-4}$, with a batch size of 16.

\subsection{Downstream Tasks}

The object-centric representation is used for various downstream tasks by training task-specific output layers. We operate on the slot embeddings produced by the pretrained model, while keeping both the AST encoder and slot attention module frozen. Given an input mixture, the model outputs a fixed number of slot embeddings, each of which is treated as a candidate representation of a sound source.

\textbf{Sound event classification:}  For classification, a shared linear layer is trained on each slot embedding to produce class logits followed by a sigmoid activation, resulting in a set of predictions per input, one for each slot. Since the ordering of slots is arbitrary, during training, we use Hungarian matching on the cross-entropy (CE) loss to assign predictions to ground-truth (GT) labels. The unmatched slots are trained to predict a dedicated background class using CE loss.

\textbf{Sound Event Detection:} For this task, a shared linear layer with sigmoid is trained on each slot embedding to produce class logits, along with a second shared linear head with sigmoid that predicts a frame-wise activity sequence for each slot. This yields per-slot temporal predictions at a fixed resolution of 100 ms, resulting in L=50 frame-level outputs for 5-second input clips, where each slot models the temporal presence of its corresponding sound event over time.

We again use Hungarian matching to assign predictions to GT events. The matching cost is defined as a combination of classification and frame-level prediction errors, computed using CE loss for class predictions and BCE loss for frame-level predictions. The final loss consists of three components: (i) classification loss for matched slot–event pairs, (ii) frame-level  prediction loss for the corresponding temporal activity, and (iii) a background classification loss for unmatched slots, which are encouraged to predict a designated background class. This formulation allows each slot to capture both the identity and temporal extent of a sound event.

\textbf{Source Separation:} In this task, each slot is used to predict a time-frequency (T-F) mask corresponding to a potential source in the mixture. 
The $D$ dimensional slot representations are first reshaped into a $D\times1\times1$ feature map and then progressively upsampled through multiple transposed convolution layers with batch normalization and ReLU activations to produce 2D spectrogram masks. The predicted masks are then resized to match the mixture spectrogram resolution using bilinear interpolation.
The predicted masks are then point-wise multiplied to the mixture magnitude spectrogram to obtain separated source estimates. The corresponding waveforms are then reconstructed using the mixture phase and the inverse short-time Fourier transform. We use Hungarian matching to align reconstructed and reference waveforms by maximizing the scale-invariant signal-to-distortion ratio (SI-SDR). The model is trained by minimizing the negative SI-SDR over the matched pairs, encouraging each slot to reconstruct a distinct source. Slots that are not assigned to any GT source are implicitly driven to represent background through the softmax operation in the model.

\section{EVALUATION} \label{evaluation}

\subsection{Dataset}\label{dataset}

We use Scaper \cite{salamon2017scaper} to construct synthetic audio mixtures using the NIGENS dataset \cite{trowitzsch2019nigens} to enable controlled training and evaluation. The NIGENS dataset contains 14 distinct sound classes - alarm, crying baby, crash, barking dog, running engine, burning fire, footsteps, knocking on door, female and male speech, female and male scream, ringing phone, piano. Mixtures are generated by randomly sampling three distinct sound classes and selecting one audio file from each class. Each audio file is resampled to 16~kHz. From each selected file, a random segment of duration between 0.5~s and 3.0~s is extracted. These segments are then placed at random temporal positions within a fixed-length mixture of 5~seconds. To simulate varying source prominence, one of the three selected classes within the mixture is randomly chosen as a reference signal. The remaining sources are scaled relative to this reference using a random gain sampled uniformly from $[-6, 6]$~dB. To prevent clipping, each mixture is peak-normalized. Along with each mixture, we store metadata including source labels, onset times, durations, and relative gain values. Using this procedure, we generate 100 hours of training data and 20 hours of test data, both consisting of 5-second mixtures. The training and test sets are constructed using audio files from different source recordings to ensure no overlap between splits.

\begin{table*}[t]
\centering
\caption{In-domain results on the NIGENS test mixtures on classification, detection, and source separation tasks.}
\label{tab:nigens_results}
\begin{tabular}{c|ccc|ccc|cc}
\hline
\textbf{Representation} & 
\multicolumn{3}{c|}{\textbf{Classification}} & 
\multicolumn{3}{c|}{\textbf{Detection}} & 
\multicolumn{2}{c}{\textbf{Separation}} \\
 & \textbf{Precision} & \textbf{Recall} & \textbf{F1} & \textbf{Precision} & \textbf{Recall} & \textbf{F1} & \textbf{SI-SDR (dB)} & \textbf{SI-SDRi (dB)} \\
\hline
Slot representation (Ours) & 0.79 & 0.76 & 0.77 & 0.78 & 0.62 & 0.66 & 10.89 & 18.41 \\
\hline
CLAP embedding & 0.75 & 0.66 & 0.68 & 0.56 & 0.13 & 0.19 & -4.17 & 3.59 \\
\hline
AST global embedding & 0.75 & 0.74 & 0.72  & 0.63 & 0.31 & 0.39 & -1.95 & 5.80 \\
\hline
AST patch embeddings & 0.77 & 0.75 & 0.74 & 0.80 & 0.58 & 0.65 & 8.14 & 15.91 \\
\hline
\end{tabular}
\end{table*}

\subsection{In-domain Evaluation on NIGENS}

For in-domain evaluation, we assess the learned slot representations on three downstream tasks: classification, detection, and separation on the NIGENS test split. These experiments evaluate whether the learned  slot representation can support downstream tasks demanding multiple granularity. All the downstream models are trained using the Adam optimizer with a learning rate of $1\times10^{-4}$. The classification and detection linear probes are trained for 20k steps, while the transposed convolution model for separation is trained for 100k steps. The classification performance is evaluated using precision, recall, and F1-score, while detection is assessed using macro-averaged precision, recall, and F1-score. Separation performance is measured using SI-SDR and SI-SDR improvement (SI-SDRi).

\subsection{Zero-shot Source Separation}

To evaluate generalization, we construct additional test mixtures from the ESC-50 dataset \cite{piczak2015esc} using the same generation procedure as described in Section \ref{dataset}. To avoid class overlap, we exclude ESC-50 classes that are present in NIGENS and use the remaining classes to synthesize 2000 mixtures. Each mixture contains three sources with randomized temporal placement and relative scaling, ensuring consistency with the training setup. These mixtures are used exclusively for zero-shot source separation evaluation and are not seen during training. The same separation pipeline is used, where the mixture is passed through the slot model to obtain the slot embeddings and these are converted to time-frequency masks and reconstructed into waveforms. This evaluation assesses whether the learned slot representations capture transferable and object-centric structure in audio, enabling the model to separate sources from previously unseen classes and distributions.

\section{Results} \label{results}

Table~\ref{tab:nigens_results} compares the proposed slot-based representations against alternative audio representations across classification, detection and separation tasks. We compare four representation types: (i) the proposed slot representations, which are a set of 1-D object-centric embeddings without an explicit temporal dimension, (ii) CLAP embeddings \cite{elizalde2023clap}, which provide a single 1-D clip-level representation for the entire audio input, (iii) global AST embeddings obtained from the special CLS token in AST \cite{gong21b_interspeech} that summarize the full input into a single representation, and (iv) AST patch embeddings corresponding to the time-frequency patch outputs of AST encoder, which retain localized temporal and spectral information.

CLAP and AST global embeddings use linear layers with sigmoid for multi-label classification and detection. For separation, the embeddings are concatenated with target class embeddings and decoded using a transposed convolutional network similar to the slot-based framework. AST patch embeddings are classified per patch and aggregated for multi-label classification, linearly projected to frame-level predictions for detection, and upsampled with transposed convolutions for source separation. More details can be found here\footref{footnote:code}.

The proposed slot representations achieve the strongest overall performance across all downstream tasks. In particular, the slot-based model achieves the highest classification F1-score of 0.77 and detection F1-score of 0.66, while substantially outperforming all baselines in source separation with an SI-SDRi of 18.41\,dB. These results indicate that the learned slot representations capture semantically meaningful, source-specific information while also implicitly preserving the temporal structure of individual sound sources.

Global representations such as CLAP embeddings and AST CLS embeddings perform notably worse, especially for source separation and SED. They compress the entire acoustic scene into a single embedding, making it difficult to disentangle overlapping sound events or preserve source-level temporal structure. This limitation is particularly evident for source separation, where both global representations produce very low SI-SDR values.

Using AST patch embeddings improves performance compared to global embeddings, particularly for detection and separation, since patch-level representations preserve localized time-frequency information. Interestingly, slot representations outperform patch embeddings despite not explicitly preserving temporal information. Unlike patch embeddings, which act as low-level local descriptors, the slot representations organize the acoustic scene into a compact set of latent object-centric representations corresponding to individual sound sources. 

To evaluate scalability, we test the model on mixtures containing four sound sources. In this setting, the model continues to separate sources effectively, achieving an SI-SDR of 7.57 and SI-SDRi of 16.75\,dB. 


\begin{table}[h]
\centering
\caption{Zero-shot separation results on ESC-50 mixtures.}
\label{tab:zeroshot_results}
\begin{tabular}{lcc}
\hline
\textbf{Representation} & \textbf{SI-SDR} & \textbf{SI-SDRi} \\
\hline
Slot representation (Ours)  & 7.55 & 14.52 \\
\hline
\end{tabular}
\end{table}

For zero-shot source separation on ESC-50 mixtures, it can be seen from Table~\ref{tab:zeroshot_results} that the slot representations generalize to unseen classes achieving an SI-SDRi of 14.52\,dB. It should be noted that, such zero-shot separation is not straightforward for other representations, since they do not explicitly provide a mechanism for source-wise decoding. In contrast, the slot-based formulation naturally supports source-wise decoding, enabling direct application to separation in unseen domains.

\section{Conclusion} \label{conclusion}

In this work, we investigated self-supervised object-centric slot-based audio representations and evaluated their effectiveness across classification, detection, and separation tasks. The proposed model consistently outperforms clip-level representations across all tasks, indicating that slot-based modeling yields more structured representations of audio scenes. The learned slots encode source-specific, semantically meaningful information and, due to the self-supervised pretraining objective, inherently capture temporal structure. This allows the model to remain competitive with time–frequency level representations on both detection and separation. The model also shows strong generalization to unseen sources, indicating that the learned representations capture compositional structure beyond the training distribution. Overall, our results suggest that slot-based modeling provides a promising direction for unified audio scene understanding. 

\printbibliography

\end{document}